\documentclass[10pt,a4paper,twocolumn]{article}
\usepackage[english]{babel} 
\usepackage[top=2.8cm,bottom=2.8cm,left=2.cm,right=2.cm]{geometry}  
\usepackage{titlesec}
\usepackage{color,soul}
\usepackage[utf8]{inputenc} 
\usepackage[affil-it]{authblk}
\usepackage{amsmath,amsfonts,amssymb}
\usepackage{varioref}
\usepackage[colorinlistoftodos]{todonotes}
\usepackage[pdftex]{hyperref} 
\let\OLDthebibliography\thebibliography
\renewcommand\thebibliography[1]{
  \OLDthebibliography{#1}
  \setlength{\parskip}{1pt}
  \setlength{\itemsep}{1pt plus 0.3ex}
}
\usepackage[binary-units = true]{siunitx}
\usepackage{float}
\usepackage{tikz}
\usepackage{xcolor}
\usepackage{comment}
\usepackage{braket}
\usepackage{nicefrac}
\usepackage{caption}
\usepackage{graphicx}  
\graphicspath{{./Figures/}{./PSTricks/}}  
\usepackage{multirow, makecell}
\usepackage{appendix}
\usepackage{setspace}
\usepackage{cite}
\usepackage{csquotes}
\usepackage{array,booktabs}

\usepackage{enumitem}
\usepackage{caption3} 
\DeclareCaptionOption{parskip}[]{}
\usepackage[font=small,labelfont=bf]{caption}
\usepackage[final]{pdfpages}
\usepackage[colorinlistoftodos]{todonotes}
\usepackage{tikz}
\usepackage{abstract}

\newcommand{\red}{\color{black}}
\newcommand{\blue}{\color{black}}
\usepackage[export]{adjustbox}
\usepackage{dblfloatfix}

\begin{document}
\title{\textbf{Efficient time-bin encoding for practical high-dimensional quantum key distribution}}
\author[1,2]{\small Ilaria Vagniluca}
\author[3]{\small Beatrice Da Lio}
\author[4]{\small Davide Rusca}
\author[3]{\small Daniele Cozzolino}
\author[3]{\small Yunhong Ding}
\author[4]{\\ \small Hugo Zbinden}
\author[1,5]{\small Alessandro Zavatta}
\author[3]{\small Leif K. Oxenløwe}
\author[3,*]{\small Davide Bacco}
\affil[1]{\footnotesize Istituto Nazionale di Ottica (INO-CNR), Largo E. Fermi 6, 50125 Florence, IT}
\affil[2]{\footnotesize Department of Physics “Ettore Pancini", University of Naples “Federico II", Via Cinthia 21, 80126 Naples, IT}
\affil[3]{\footnotesize CoE SPOC, DTU Fotonik, Technical University of Denmark, 2800 Kgs. Lyngby, DK}
\affil[4]{\footnotesize Group of Applied Physics, Université de Genève, Chemin de Pinchat 22, 1211 Geneva 4, CH}
\affil[5]{\footnotesize LENS and Department of Physics, University of Florence, 50019 Sesto Fiorentino, IT}
\affil[*]{dabac@fotonik.dtu.dk}

\date{} 
\pagestyle{plain}
\setcounter{page}{1}
\twocolumn[ 
\begin{@twocolumnfalse}
\maketitle
     \vspace{-0.8cm}
\begin{abstract}
\normalsize
\vspace*{-1.0em}
\noindent
High-dimensional quantum key distribution (QKD) allows to achieve information-theoretic secure communications, providing high key generation rates which cannot in principle be obtained by QKD protocols with binary encoding. Nonetheless, the amount of experimental resources needed increases as the quantum states to be detected belong to a larger Hilbert space, thus raising the costs of practical high-dimensional systems. Here, we present a novel scheme for fiber-based 4-dimensional QKD, with time and phase encoding and one-decoy state technique. Quantum states transmission is tested over different channel lengths up to \SI{145}{\kilo\metre} of standard single-mode fiber, evaluating the enhancement of the secret key rate in comparison to the three-state 2-dimensional BB84 protocol, which is tested with the same experimental setup. Our scheme allows to measure the 4-dimensional states with a simplified and compact receiver, where only two single-photon detectors are necessary, thus making it a cost-effective solution for practical and fiber-based QKD.

\vspace{0.5cm}
\end{abstract}
  \end{@twocolumnfalse}
 ]

\section*{Introduction}
\vspace{-0.25cm}
As the constant advancement in quantum computing is threatening the security of current cryptographic systems, our society needs an alternative technology to safely transmit sensitive data and confidential information~\cite{Nist2016}. A quantum-proof solution to safely deliver our cryptographic keys is quantum key distribution (QKD), which exploits quantum states of light as safeguarded bit carriers over untrusted communication channels{\red~\cite{BB84,Gisin2002,Pirandola2019,xu2019quantum}}. In well-established QKD protocols such as the BB84~\cite{BB84}, each bit of the key is carried by a single photon, which is prepared in order to span a set of different quantum states belonging to a 2-dimensional Hilbert space, \textit{i.e.} qubits. High-dimensional QKD protocols were introduced more recently~\cite{Bechmann2000,Cerf2002}, proving how $n=\log_2(d)>1$ bits of information can be safely encrypted on each single photon, by preparing an enlarged set of states belonging to a $d$-dimensional Hilbert space. Such states are called qudits. 
{\red The higher information capacity of qudits allows for an optimized exploitation of the photon budget at the transmitter; at the same time it also mitigates the issue of saturation in the receiver's single-photon detectors. Moreover, using high-dimensional states improves the robustness to the noise affecting the communication, allowing for a higher threshold value of the quantum bit error rate (QBER).} The result is an increase in the secret key rate achievable by high-dimensional QKD, as compared with standard QKD protocols with binary encoding ($d=2$), at least until the overall losses are low enough to keep negligible the random dark counts at the receiver~\cite{Sheridan2010,cozzolino2019review,bacco2019}.\\
Although there are many degrees of freedom to be exploited to send more than one bit per photon~\cite{Groblacher2006,mirhosseini2015,canas2017,ding2017,cozzolino2019orbital}, time-bin and time-energy encoding are the ones more suitable for single-mode fiber propagation, and thus more compatible with the already existing and widespread fiber networks~\cite{zhong2015,Lee2016,bacco2016,Islam2017,ikuta2018,Boaron2018_421km,dalio2019,Islam2019}. Recent demonstrations of one-way fiber-based QKD include the record-breaking key rate of \SI{26.2}{\mega\bit\per\second} at \SI{4}{\decibel} channel loss~\cite{Islam2017}, achieved with a 4-dimensional time-bin protocol with two decoy states, which is proven to be robust against the most general (or coherent) attacks as well as finite-size effects. 
However, the apparent gain in the key rate comes with a cost, as the preparation and measurement of high-dimensional states require a larger amount of experimental resources, especially at the receiver, who has to project the incoming qudits on two unbiased bases of $d$ orthogonal states. For instance, to perform the time-bin protocol presented in~\cite{Islam2017}, at least three cascaded interferometers and five single-photon detectors are necessary to measure the 4-dimensional states (actually three more detectors were added in the cited work, in order to reduce saturation effects in the first basis measurements). On the other hand, a simpler receiver with only two single-photon detectors is sufficient in many of the binary-encoded protocols with the same level of security, such as the BB84~\cite{ruiz2011,Boaron2018_421km,grunenfelder2018}.\\
In this work we present an alternative scheme for 4-dimensional QKD with time-bin encoding, which allows to implement a simplified and more compact receiver, where only two single-photon detectors are necessary for measuring all the quantum states. The security of this protocol against general attacks is demonstrated in a finite-key scenario, when a simple and efficient one-decoy state method is implemented. Qudits exchange was tested over different fiber channels up to \SI{145}{\kilo\metre} of length, corresponding to \SI{31.5}{\decibel} of transmission loss. In addition to this, in order to evaluate the improved performances of our protocol, we tested also a 2-dimensional BB84 scheme over the same channel lengths, by employing mostly the same experimental setup. {\red To be noted that both QKD systems employ two single-photon detectors, \textit{i.e.}, the new proposal is cost-effective.} This method allowed us to make a fair and rational comparison between the two time-bin protocols with $d=2$ and $d=4$. In the following sections, we describe the two protocols and we report the security analysis of our new scheme. We then show the experimental setup of the transmitter and the receiver. Finally, our results are presented and discussed in the last section of the paper.

\section*{Protocols}
\label{sec:protocols}
Figure~\ref{fig:1} schematically depicts the quantum states and mutually unbiased bases belonging to the two different QKD schemes that were performed in this work. The 2- and 4-dimensional protocols implemented are the three-state time-bin BB84 and one of its possible generalizations in 4 dimensions, respectively. Both schemes are secure against general eavesdropping attacks that are addressed to the transmission channel, as discussed later for a finite-key analysis. Defining $\tau$ as the bin duration, each qubit and qudit has a time span of two and four bins respectively. In both cases, quantum states of the $\mathcal{Z}$~basis are adopted for key bits encoding, while the $\mathcal{X}$~basis is implemented only for security checking.\\ 
In the three-state BB84~\cite{Boaron2018_421km}, quantum states belonging to the $\mathcal{Z}$~basis differ for the time bin occupied by the photon and only one bit, corresponding to early or late bin occupation, is encoded on each state. The third state is the superposition of the two $\mathcal{Z}$~basis states with $0$ relative phase, while the other orthogonal state in the $\mathcal{X}$~basis (with $\pi$ relative phase) is not prepared, thus making this protocol a simplified version of the original four-state BB84. Here, the projection on the $\mathcal{Z}$~basis is made at the receiver by measuring the photon arrival time with a single-photon detector, while the $\mathcal{X}$~basis is measured by monitoring, with another single-photon detector, one of the two outputs of a Mach-Zehnder interferometer with a delay equal to $\tau$. Whenever weak coherent pulses are prepared instead of single-photons (as in our case), an efficient one-decoy scheme can be implemented in order to avoid photon number splitting attacks~\cite{lo2005,Rusca2018_FiniteKeyAnalysis}. The secret key length $ \ell_{2D} $ per privacy amplification block is then given by the following formula:
\begin{figure}[t]
\centering
\includegraphics[width=0.4\textwidth]{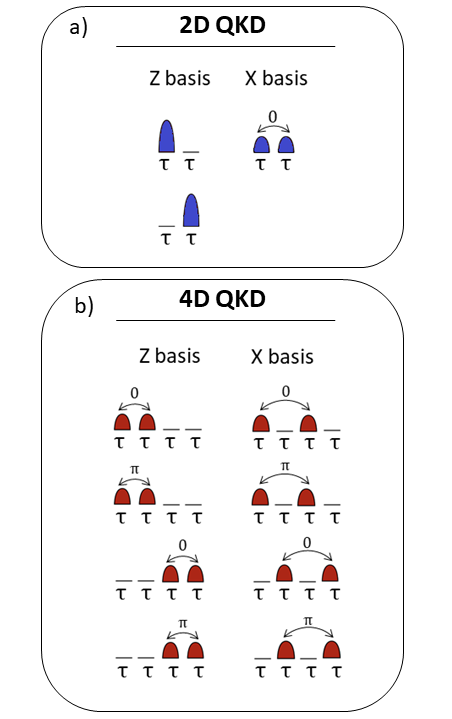}
\caption{{\bf Quantum states involved in the two QKD protocols.} The three states belonging to the 2-dimensional BB84 (a) and the eight states belonging to the novel 4-dimensional protocol (b). $\tau$ is the time bin duration, while $0$ and $\pi$ specify the relative phase between the different time bins occupied by the photon.}
\label{fig:1}
\end{figure}
\begin{equation}
\begin{split}
    \ell_{2D} \ \ \leq \ \  & D_0^{\mathcal{Z}} \ \ + \ \ D_1^{\mathcal{Z}} \Bigl[ 1-h\bigl( \phi_{\mathcal{Z}} \bigr) \Bigr] \ \ - \ \ \lambda_{EC} \ \\
    & \ \ - \ \ 6\log_2(19/\epsilon_{sec}) \ \ - \ \ \log_2(2/\epsilon_{corr})
\end{split}
\end{equation}
where $D_0^{\mathcal{Z}}$ and $D_1^{\mathcal{Z}}$ are the lower bounds of vacuum events and single-photon events in the $\mathcal{Z}$~basis, $h(\cdot)$ is the binary entropy function, $\phi_{\mathcal{Z}}$ is the upper bound on the phase error rate and $\lambda_{EC}$ is the number of bits that are publicly announced during error correction, while $\epsilon_{sec}$ and $\epsilon_{corr}$ are the secrecy and correctness parameters~\cite{Rusca2018_SecProofSimpleBB84}.\\
The direct generalization in 4 dimensions of the time-bin BB84 is the protocol presented in~\cite{Islam2017}, where the four qudits belonging to the $\mathcal{Z}$~basis differ only for the time bin which is individually occupied among the four bins available, and each state of the $\mathcal{X}$~basis is a superposition of all the four bins, combined with different relative phases. In that case, the projection on the $\mathcal{Z}$~basis is made simply with a detector measuring the time of arrival of the photons, as for the 2-dimensional protocol. On the other hand, the projection on the mutually unbiased $\mathcal{X}$~basis requires a more complicated setup, with at least a cascade of three interferometers and four detectors (as reported in~\cite{Islam2017} and~\cite{Islam2017interf}) or even more complex solutions (as discussed in~\cite{Brougham2013}).\\
In our new scheme we exploit two more convenient 4-dimensional bases, that are depicted in Figure~\ref{fig:1}b. Here, we have two time bins combined for each state in both bases, and the four states that are defined on consecutive bins are employed for key encoding in the $\mathcal{Z}$~basis. Qudits belonging to the same basis are orthogonal to each other and the two bases are mutually unbiased, since the general relation
\begin{equation}
    \Bigl\lvert \braket{z_n|x_m}\Bigr\rvert^2=\frac{1}{d}
\end{equation}
is still satisfied for all $\ket{z_n}$ and $\ket{x_m}$ states belonging to $\mathcal{Z}$ and $\mathcal{X}$~basis respectively ($n,m=0,...,d-1$). For this scheme, the projection on the $\mathcal{Z}$~basis is made with one single $\tau$-delayed interferometer, while the projection on the $\mathcal{X}$~basis requires a single $2\tau$-delayed interferometer. This makes it possible to hugely simplify the experimental setup at the receiver side, in comparison to ref.~\cite{Islam2017}, as it is further described in the following section.\\
The secret key length $ \ell_{4D} $ per privacy amplification block is given by:
\begin{equation}
\begin{split}
    \ell_{4D} \ \ \leq \ \  & 2D_0^{\mathcal{Z}} \ \ + \ \ D_1^{\mathcal{Z}} \Bigl[ 2-H\bigl( \phi_{\mathcal{Z}} \bigr) \Bigr] \ \ - \ \ \lambda_{EC} \ \\
    & \ \ - \ \  6\log_2(19/\epsilon_{sec}) \ \ - \ \ \log_2(2/\epsilon_{corr})
\end{split}
\end{equation}
where $H(x) := - x \ \log_2(x/3) - (1 - x) \ \log_2(1-x)$ is the Shannon entropy in a 4-dimensional Hilbert space. The lower and upper bounds on the single-photon events in the equation above are obtained by using the one-decoy technique appearing in ref.~\cite{Rusca2018_FiniteKeyAnalysis}, modified for the 4-dimensional QKD protocol. 
The difference between the original one-decoy protocol~\cite{Rusca2018_FiniteKeyAnalysis} and the one presented here is the method to find the upper bound to the vacuum events, $D_0^{\mathcal{Z},u}$. In the 4-dimensional QKD each basis measurement has four possible outputs, meaning that the probability of error due to a vacuum event is $3/4$. By exploiting this fact, the vacuum events can be bounded by the total number of errors $m_{\mathcal{Z},k}$ in the $\mathcal{Z}$~basis corresponding to the decoy intensity $k$. By correcting the estimated quantities using the finite-key technique presented in ref.~\cite{Rusca2018_FiniteKeyAnalysis}, the upper bound to the vacuum events is given by the following expression:

\begin{multline}
D_0^{\mathcal{Z}} \leq D_0^{\mathcal{Z},u} := \frac{4}{3}\Biggl[\tau_0\frac{e^k}{p_k}\Bigl(m_{\mathcal{Z},k} + \sqrt{\frac{m_{\mathcal{Z}}}{2}\log\frac{1}{\varepsilon_2}} \ \Bigr) \\
 + \sqrt{\frac{n_{\mathcal{Z}}}{2}\log\frac{1}{\varepsilon_1}} \ \Biggr]
\end{multline}
where $\tau_0 = \sum_{k\in\kappa}p_ke^{-k}$ is the total probability to send the vacuum state, $p_k$ is the probability to prepare the decoy state of intensity $k$, $n_{\mathcal{Z}}$ and $m_{\mathcal{Z}}$ are, respectively, the total number of detections and the total number of errors in the $\mathcal{Z}$~basis.

\section*{Experimental setup}
\label{sec:setup}
\begin{figure*}[t]
\centering
\includegraphics[width=1\textwidth]{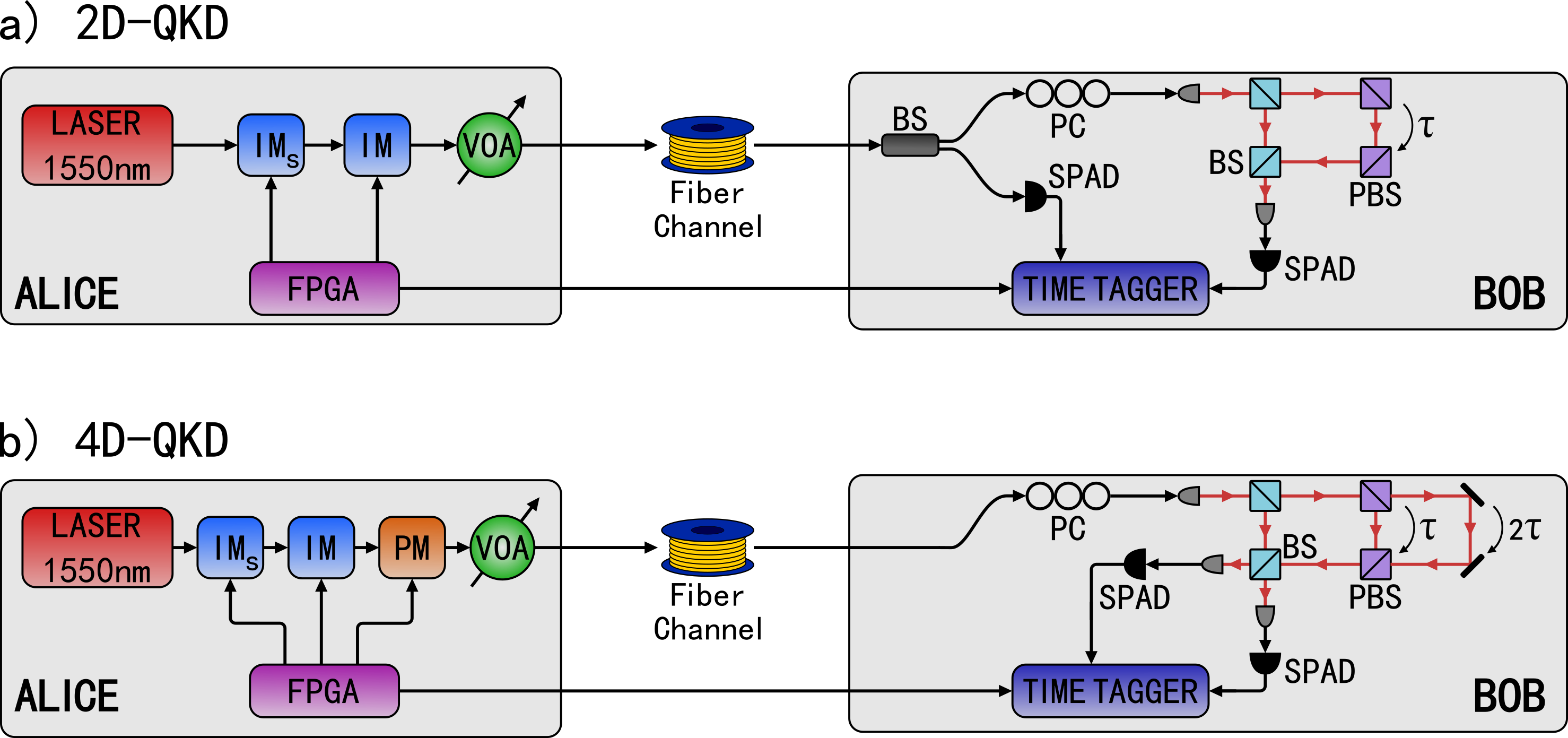}
\caption{{\bf Experimental setup for the two QKD schemes.} Transmitter (Alice) and receiver (Bob) employed to test the three-state BB84 (a) and the 4-dimensional protocol (b). Black lines represent fiber optic and electrical cables while red lines stand for free-space propagation. The same setup was employed to test both protocols, thus the three-state BB84 was performed by using one of the two overlapped interferometers that are shown in (b). IM: intensity modulator, PM: phase modulator, VOA: variable optical attenuator, FPGA: Field Programmable Gate Array, PC: polarization controller, BS: beam splitter, PBS: polarizing beam splitter, SPAD: single-photon avalanche detector. $\tau$ and $2\tau$ are the delays in time corresponding to the two overlapped interferometers at the receiver side.}
\label{fig:2}
\end{figure*}
\vspace{-0.25cm}
The experimental setup of the two QKD schemes performed is illustrated in Figure~\ref{fig:2}. The transmitter (Alice) is very similar in both cases: a train of weak coherent pulses is prepared from a continuous-wave (CW) laser emitting at \SI{1550}{\nano\metre}, by means of sequential intensity modulators (IM). We employed two cascaded IMs in order to optimize the pulse carving, while a third one is used for implementing the one-decoy state technique. A phase modulator (PM) modulates the relative phase between the time bins, necessary for qudits preparation. Finally, a variable optical attenuator (VOA) is used to reach the single-photon level before sending the pulses into the fiber channel. 
All the optical modulators at the transmitter side are driven with a field programmable gate array (FPGA). For carving the CW laser we used a custom sequence of electrical pulses (of about \SI{150}{\pico\second} of width), which already includes Alice's state and basis choice. The time bin duration is $\tau=\SI{840}{\pico\second}$, resulting in a qubit rate of about \SI{595}{\mega\hertz} and a qudit rate of approximately \SI{297.5}{\mega\hertz}. A pseudo random binary sequence (PRBS) of $2^{12}-1$ symbols and a symbol width of $d\cdot\tau$ (with $d=2$ or $4$ depending on the protocol) is used to drive the third IM, in order to send the two different intensity levels corresponding to signal ($\mu_1$) and decoy ($\mu_2$) states. With this configuration, the probability to send $\mu_1$ or $\mu_2$ is fixed to 50\% (for both qubits and qudits preparation) and can not be optimized for each different channel loss. To prepare the qudits, another PRBS of $2^{12}-1$ symbols and a symbol width equal to $\tau$ is used to drive the PM. Please notice that the phase randomization of quantum states (required to guarantee the security of the decoy-state method) can be easily performed with another phase modulator, or by employing a pulsed laser source working in gain-switching mode \cite{Boaron2018_Simple2.5GHzQKD,Boaron2018_421km}. \\
At the receiver side (Bob) two overlapped free-space interferometers were installed, as shown in Figure~\ref{fig:2}b. The short arm is in common, while the long arm is selected between two different paths (with delay equal to $\tau$ and $2\tau$) by means of two polarizing beam splitters (PBS). 
{\blue Light with vertical polarization is reflected by the PBS and follows the $\tau$ delay-line, while horizontally polarized light is transmitted by the PBS and follows the $2\tau$ delay-line.} To perform our measurements, the polarization of {\blue quantum pulses entering Bob's station was manually adjusted} with a fiber-based polarization controller (PC). The overall loss of the $\tau$-delayed and $2\tau$-delayed interferometer is \SI{2.3}{\decibel} and \SI{2.5}{\decibel} respectively, due to imperfect beam splitting and fiber coupling at the detectors. 
Additional losses of about \SI{9.2}{\decibel} are due to the detection efficiency ($20\%$) and timing resolution (\SI{200}{\pico\second}) of the single-photon avalanche detectors (SPADs). The detectors dead time is \SI{20}{\micro\second}, therefore their click rate saturates when it approaches the value of \SI{50}{\kilo\hertz}.\\
To perform the 2-dimensional protocol, the receiver passively selects his basis with a beam splitter (BS), as it is shown in Figure~\ref{fig:2}a. One SPAD measures the time of arrival while the other SPAD monitors an output of the $\tau$-delayed interferometer. A time tagging unit, which is synchronized with Alice's FPGA via a classical channel, collects the electrical outputs from the two SPADs and finally transmits the acquired data to Bob's computer.  
The polarization of free-space light is kept aligned with the vertical direction while testing the three-state BB84. Please notice that, to perform this protocol, the two PBS can be replaced with two standard mirrors, which make unnecessary the polarization controlling. However, here both protocols were tested with the same experimental setup, thus one of the two overlapped interferometers of the 4-dimensional scheme (shown in Figure~\ref{fig:2}b) was employed to perform also the three-state BB84. \\
In order to test the 4-dimensional protocol, both outputs of the interferometers were monitored with the two SPADs, and basis selection at the receiver is made by manually switching the polarization between the two directions (Figure~\ref{fig:2}b). This means that only one of the two 4-dimensional bases was prepared and measured at a time and therefore, no real-time basis choice was performed for the 4-dimensional protocol during this experiment (on the other hand, the three-state BB84 was tested with real time basis choice at both the transmitter and the receiver sides). The interference of 4-dimensional states is observed in the second and fourth time bins for the $\mathcal{Z}$~basis measurements, and in the third and fourth time bins for the $\mathcal{X}$~basis measurements.  
{\blue The receiver can uniquely determine the output of his projection by observing the time bin and the detector at which the click occurs.} 
{\red To be noted that, even though the observed time bins at the detection output are correlated with the polarization of incoming light, this does not represent an advantage for the eavesdropper, who is still unable to control the detection efficiency in Bob's measurement bases without being noticed. Anyway, adding a polarizer in front of Bob's setup would definitely filter out any component of residual light in the wrong polarization.}


\section*{Results and discussion}
\begin{table*}[t]
\setlength{\tabcolsep}{14pt}
\caption{{\bf Experimental parameters and results.} Here are reported the values that we set at the transmitter and the receiver for each fiber channel, such as the mean photon number for signal and decoy states ($\mu_1$, $\mu_2$) and the probabilities to prepare and measure the $\mathcal{Z}$~basis ($p_{\mathcal{Z}}^{Alice}$, $p_{\mathcal{Z}}^{Bob}$). From the acquired data we measured the quantum bit error rate (QBER) and the upper bound on the phase error rate ($\phi_{\mathcal{Z}}$) in the $\mathcal{Z}$~basis; then we evaluated the final secret key rate (SKR). The block size is fixed to $10^7$ for all channel lengths. The state preparation rate ($\textrm{R}$) is \SI{595}{\mega\hertz} for qubits and \SI{297.5}{\mega\hertz} for qudits.}
\begin{center}
\begin{tabular}{|c|c|cccc|}
    \hline\hline
    &&&&& \\[-0.4em] 
    \multirowcell{2}{{\bf transmission}\\{\bf channel}}
    & length & \SI{25}{\kilo\meter} & \SI{65}{\kilo\meter} & \SI{105}{\kilo\meter} & \SI{145}{\kilo\meter} \\[7pt]
    & loss & \SI{5.1}{\decibel} & \SI{14}{\decibel} & \SI{23}{\decibel} & \SI{31.5}{\decibel} \\[7pt]
    \hline
    &&&&& \\[-0.6em] 
    \multirowcell{7}{{\bf 2-dimensional} \\{\bf three-state BB84} \\ {\bf protocol} \\ \\
    {($\textrm{R}=\SI{595}{\mega\hertz}$)}}
    & $\mu_1$ & 0.07 & 0.12 & 0.26 & 0.31 \\[7pt]
    & $\mu_2$ & 0.03 & 0.06 & 0.14 & 0.15 \\[7pt]
    & $p_{\mathcal{Z}}^{Alice}$ & 0.9 & 0.9 & 0.9 & 0.9 \\[7pt]     
    & $p_{\mathcal{Z}}^{Bob}$ & 0.5 & 0.9 & 0.5 & 0.5 \\[7pt]    
    & QBER & 1.1\% & 1.1\% & 1.4\% & 2.3\% \\[7pt]     
    & $\phi_{\mathcal{Z}}$ & 6.6\% & 9.2\% & 8.9\% & 13.6\% \\[7pt]   
    & SKR & \SI{15}{\kilo\bit\per\second} & \SI{12}{\kilo\bit\per\second} & \SI{5.1}{\kilo\bit\per\second} & \SI{0.53}{\kilo\bit\per\second} \\[7pt]    
    & SKR/R & \SI{2.6e-5}{} & \SI{2.0e-5}{} & \SI{8.7e-6}{} & \SI{8.9e-7}{} \\[7pt]    
    \hline
    &&&&& \\[-0.6em] 
    \multirowcell{7}{{\bf 4-dimensional} \\ {\bf protocol} \\  \\ {($\textrm{R}=\SI{297.5}{\mega\hertz}$)}}
    & $\mu_1$ & 0.10 & 0.20 & 0.21 & 0.18 \\[7pt]
    & $\mu_2$ & 0.05 & 0.10 & 0.10 & 0.08 \\[7pt]
    & $p_{\mathcal{Z}}^{Alice}$ & 0.9 & 0.9 & 0.9 & 0.9 \\[7pt]    
    & $p_{\mathcal{Z}}^{Bob}$ & 0.7 & 0.7 & 0.7 & 0.5 \\[7pt]    
    & QBER & 3.4\% & 3.4\% & 4.9\% & 7.9\% \\[7pt]     
    & $\phi_{\mathcal{Z}}$ & 3.9\% & 4.6\% & 5.7\% & 7.2\% \\[7pt]   
    & SKR & \SI{37}{\kilo\bit\per\second} & \SI{24}{\kilo\bit\per\second} & \SI{5.5}{\kilo\bit\per\second} & \SI{0.42}{\kilo\bit\per\second} \\[7pt]    
    & SKR/R & \SI{1.2e-4}{} & \SI{7.9e-5}{} & \SI{1.8e-5}{} & \SI{1.4e-6}{} \\[7pt] 
    \hline\hline
\end{tabular}
\end{center}
\label{tab:1}
\end{table*}
\vspace{-0.3cm}
\begin{figure}[t]
\centering
\includegraphics[width=0.53\textwidth]{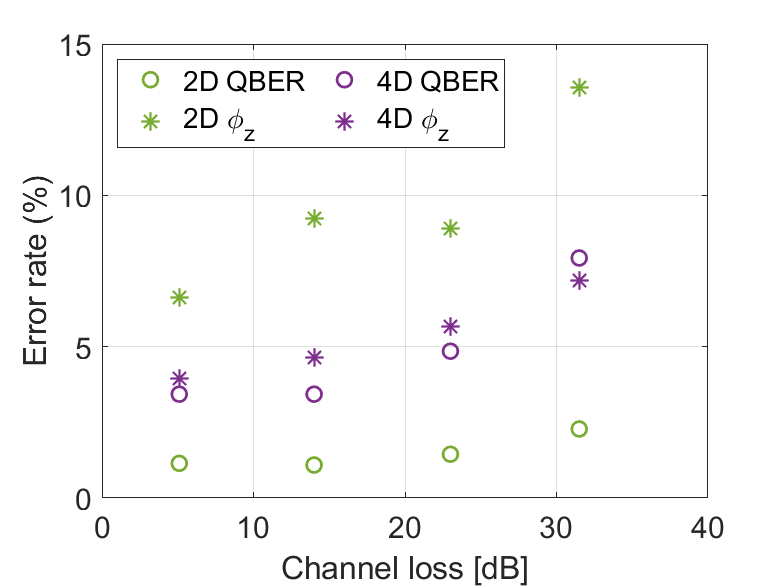}
\caption{{\bf Error rates measured for each protocol.} Quantum bit error rate (QBER) and upper bound on the phase error rate ($\phi_{\mathcal{Z}}$) experimentally measured for the two protocols, at the different channel losses.}
\label{fig:qber}
\end{figure}
\begin{figure}[t]
\centering
\includegraphics[width=0.53\textwidth]{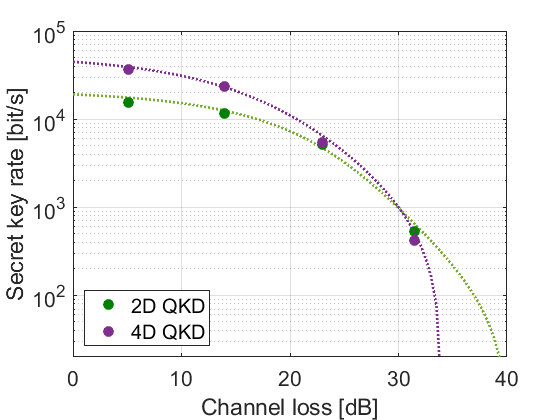}
\caption{{\bf Secret key rate as a function of channel loss.} Each point represents the secret key rate evaluated from the experimental data. Dashed lines reproduce the simulated behaviour of the secret key rate achievable by our setup, for the two QKD protocols.}
\label{fig:skr}
\end{figure}
The two QKD schemes were tested over different channel lengths of standard single-mode fiber. The experimental parameters and results are summarized in Table~\ref{tab:1}. For each transmission channel we experimentally set, at the transmitter side, the optimal values for the mean photon number of signal and decoy states ($\mu_1$ and $\mu_2$), that were previously estimated in order to maximize the secret key rate (SKR) achievable by each protocol. As already mentioned, the probability to send $\mu_1$ or $\mu_2$ is fixed to $50\%$ in our experimental setup. \\
For the three-state BB84 we used a pulse sequence on the FPGA consisting of $90\%$ of $\mathcal{Z}$~basis states, thus the basis choice at the transmitter is $p_{\mathcal{Z}}^{Alice}=0.9$ for all channel lengths. At the receiver side we set $p_{\mathcal{Z}}^{Bob}$ equal to $0.5$ or $0.9$ (see Table~\ref{tab:1}), depending on the splitting ratio of the BS that was selected in order to maximize the SKR. For the 4-dimensional protocol we tested only one basis at a time, thus the probabilities $p_{\mathcal{Z}}^{Alice} , \ p_{\mathcal{Z}}^{Bob}$ were numerically set during the evaluation of the final SKR. We fixed $p_{\mathcal{Z}}^{Alice}=0.9$ as for the 2-dimensional protocol, and again we selected $p_{\mathcal{Z}}^{Bob}$ from two different values ($0.7$ {\blue or $0.5$, see Table~\ref{tab:1}}) in order to get the highest SKR at each channel length. \\ 
From the acquired data we evaluated the quantum bit error rate in the $\mathcal{Z}$~basis (QBER) and in the $\mathcal{X}$~basis, which gives the upper bound on the phase error rate ($\phi_{\mathcal{Z}}$). {\blue Notice that with this terminology (the same adopted in most of the previous works) we always refer to the symbol-error rate, which exactly matches the bit-error rate only in the 2-dimensional case.} These data are plotted in Figure~\ref{fig:qber} for both protocols. As expected, the error rates appear to increase with the channel loss, due to the random noise counts which become more and more significant as we leave the saturation regime of the single-photon detectors. Noise counts include the detectors dark counts, the background photons entering in the fibers, and also the imperfect modulation of light pulses at the transmitter side. The QBER is lower for the 2-dimensional protocol, since measuring only the arrival time of weak pulses is generally more straightforward than measuring also their phase, which requires the interference to be maximized and stabilized. 
{\blue On the other hand, $\phi_{\mathcal{Z}}$ is lower in the 4-dimensional case, since here the photons are collected simultaneously from both outputs of each interferomenter. This configuration, where both outputs are monitored at the same time, practically results in a more stable optimization of interference during the acquisition. As a consequence, the measurement of relative phase exhibits less noise, as compared with the 2-dimensional protocol, where only a single output of the interferometer is monitored during the acquisition.}
Moreover, since the measurement method of the two bases is the same in the 4-dimensional case {\blue (involving both phase and time simultaneously)}, the values of the QBER and $\phi_{\mathcal{Z}}$ are more similar to each other, in comparison to the error rates of the 2-dimensional protocol, where each basis is measured in a different way {\blue (involving only phase or only time separately)}. The value of $\phi_{\mathcal{Z}}$ also depends on the total amount of detections in each basis and is affected by the different setting of $p_{\mathcal{Z}}^{Bob}$.\\
From Bob's detection data we computed the final SKR achievable in a finite-key scenario, by setting a block size of $n_{\mathcal{Z}}=10^7$ in the $\mathcal{Z}$~basis, and a secrecy and correctness parameters of $10^{-9}$. In Table~\ref{tab:1} is reported also the secret fraction SKR/R, which estimates how many secret key bits can be extracted from each quantum state that is initially prepared. 
Our results, which are plotted in Figure~\ref{fig:skr}, show an enhancement of the SKR achievable by the 4-dimensional protocol for the first two experimental points, for which the SKR is increased by a factor $2.4$ and $2.0$ respectively. For higher channel loss, the SKR decreases faster than in the three-state protocol, in agreement with the expected behaviour which is represented by the dashed lines in Figure~\ref{fig:skr}. Indeed, our experimental setup allows to extract a secret key up to \SI{39}{\decibel} channel loss with the three-state BB84, and up to \SI{34}{\decibel} with the 4-dimensional protocol. This is due to the fact that a random noise count has $1-1/d$ probability to generate an error at the receiver, where $d$ is the dimension of the encoding: the higher the Hilbert space dimension, the more effective is the random noise at the receiver. On the other hand, the doubled information capacity and the enhanced resilience to state perturbations, make it possible to increase the SKR by more than a factor $2$ in the saturation regime of the single-photon detectors. In addition, the secret fraction SKR/R is improved by the 4-dimensional protocol for all the experimental points (as shown in Table~\ref{tab:1}){\red: less photons are necessary to deliver the same secret key.} This means that if we prepare the qudits at the same rate as used for the qubits, we can increase the SKR for all of the four channel lengths.


\section*{Conclusions}
In conclusion, we have presented a fiber-based 4-dimensional QKD protocol with an efficient time and phase encoding scheme, which has the advantage to require a very simple and compact setup at the receiver. This novel QKD scheme was experimentally tested over different channel lengths, and its performances were compared with the three-state BB84, a well-established 2-dimensional protocol which was also tested in this work. 
{\red Mostly the same experimental setup was employed to test the two protocols, including the same amount of single-photon detectors at the receiver, as well as the same time-bin duration (which resulted in a halved preparation rate of 4-dimensional states at the transmitter). In this configuration,} we demonstrated an enhancement of the secret key rate by a factor $2.4$ in the saturation regime of the detectors, by testing only one 4-dimensional basis at a time. In the future, we plan to perform a real-time basis choice at the receiver, by adding a polarization switcher (an off-the shelf component for fiber-based telecommunications). This new device will introduce an extra loss at the receiver (of about \SI{2}{\decibel}), which in any way is low enough to not affect the improved performances of our 4-dimensional scheme in the saturation regime. Moreover, the effect of this extra loss can be easily balanced by reducing the other sources of loss at the receiver, or by optimizing all of the experimental parameters at the transmitter side ({\red such as} the basis choice and the decoy probabilities, that were both fixed during this experiment). Furthermore, our system could be easily modified to implement the two-decoy state technique, which is more resilient to noise. This would allow us to optimize the protocol for each configuration of the experimental parameters, increasing the overall performance once more.\\
Our demonstration proves that high-dimensional quantum systems allow a notable improvement in the key generation process as compared with the binary-encoding case. At the same time, no extra expensive resources are necessary for the full implementation of such a system. Thus, our experiment paves the way towards a wider use of high-dimensional encoding in quantum communication.


\section*{Acknowledgements}
\vspace{-0.25cm}

\section*{Funding}
\vspace{-0.25cm}
This work is supported by the Center of Excellence, SPOC - Silicon Photonics for Optical Communications (ref DNRF123), by the EraNET Cofund Initiatives QuantERA within the European Union’s Horizon 2020 research and innovation program grant agreement No.
731473 (project SQUARE), by the NATO Science for Peace and Security program  under grant G5485 and by the European Union’s Horizon 2020 program under the Marie Sk\l{}odowska-Curie project QCALL (GA 675662).

\section*{Abbreviations}
QKD Quantum key distribution; BB84 Bennett and Brassard 1984 QKD protocol; 2D two-dimensional; 4D four-dimensional; IM Intensity Modulator; CW continuous wave; PM Phase Modulator; BS beam splitter; PBS polarizing beam splitter; VOA variable optical attenuator; FPGA Field Programmable Gate Array; PRBS pseudo random binary sequence; SPAD Single Photon Avalanche Detectors; QBER Quantum Bit Error Rate; SKR Secret Key Rate.



\section*{Authors’ contributions}
D.B. and I.V. conceived the experiment. I.V., B.D.L., D.C., and D.B. carried out the experimental work. D.R. and B.D.L. carried out the theoretical analysis on the protocols. All authors discussed the results and contributed to the writing of the manuscript.

\bibliography{mybib}{}
\bibliographystyle{ieeetr}

\end{document}